\documentclass[aps,prl,amsmath,amssymb,twocolumn]{revtex4-2}

\usepackage{graphicx}
\usepackage{bm,color}
\usepackage{physics}
\usepackage{amsmath, amssymb}
\usepackage{bm,color}
\usepackage{multirow}
\usepackage{ulem}
\usepackage{bbm}
\usepackage{bm}
\usepackage[colorlinks,bookmarks=true,citecolor=blue,linkcolor=red,urlcolor=blue]{hyperref}

\newcommand{\be}{\begin{eqnarray}}
\newcommand{\ee}{\end{eqnarray}}

\begin{document}

\title{Emergence symmetry protected topological phase in spatially tuned measurement-only circuit}
\date{\today}
\author{Yoshihito Kuno$^{1}$}
\author{Ikuo Ichinose$^{2}$}
\thanks{A professor emeritus}

\affiliation{$^1$Graduate School of Engineering Science, Akita University, Akita 010-8502, Japan}
\affiliation{$^2$Department of Applied Physics, Nagoya Institute of Technology, Nagoya, 466-8555, Japan}

\begin{abstract}
Topological phase transition induced by spatially-tuned single-site measurement is investigated in a measurement-only circuit, in which
three different types of projective measurement operator.
Specific spatial setting and combination of commutation relations among three measurement operators generate such a transition.
In practice, symmetry protected topological (SPT) phase is recovered on even sublattice by eliminating a projective measurement disturbing the SPT 
via applying another spatially-tuned projective measurement on odd sublattice. 
We further investigate the critical properties of the phase transition and find that it has the same critical exponents with 
the two-dimensional percolation transition.  
\end{abstract}


\maketitle
{\it Introduction.---}
Measurement of quantum many-body system induces nontrivial dynamical effects. 
One of the most interesting phenomena induced by measurements is entanglement phase transition in hybrid random unitary circuits \cite{Li2018,Skinner2019,Li2019,Vasseur2019,Chan2019,Szyniszewski2019,Choi2020,Bao2020,Jian2020,Vijay2020,Sang2021,Sang2021_v2,Nahum2021,Hashizume2022,Fisher2022_rev}. 
This phase transition phenomenon appears in various hybrid circuits including a time evolution operators of some many-body Hamiltonian \cite{Fuji2020,Goto2020,Tang2020,Lunt2020,Turkeshi2021,Kells2022,Fleckenstein2022,KOH2022}.  
High entanglement of states generated by a unitary time evolution is suppressed by the measurements. 
Also, as a typical non-equilibrium dynamical aspect, the spread of entanglement or scrambling of the quantum information is suppressed.
This change of states on the circuits is not a crossover but phase transition. 
Interestingly, the criticality of phase transitions in the circuits is mostly classified into a universality class, such as two-dimensional (2D) percolation \cite{Li2019,Zabalo2020,Sharma2022}.
Besides unitary evolution with measurements, measurement-only circuit \cite{Lang2020,Ippoliti2021} also displays striking phenomena, i.e., 
combination of multiple kinds of measurements
can induce novel phase transitions and generates non-trivial states such as measurement-only thermal state not exhibiting area law of entanglement entropy \cite{Ippoliti2021,Lavasani2021_2}, 
symmetry protected topological (SPT) state \cite{Lavasani2021,Klocke2022} and topological order \cite{Lavasani2021_2}.
It also gives various interesting critical phenomena of the measurement-induced phase transition. 
In particular, measurement-only circuit with a specific frustration graph \cite{Chapman2020,Ippoliti2021}, which
clarifies network relationship of anti-commutation relation between measurement operators, 
enhances complexity and non-triviality of resultant steady states.
Nevertheless, it exhibits universal behavior at transition points, i.e.,
various phase transitions in various circuits can be classified into a unique or closely-related universality class \cite{Li2018,Skinner2019,Li2019,Vasseur2019}. 

In general, various setups of the measurement-only circuit are possible, although only some of them have been explored so far.
As one of interesting properties of measurement-only circuits, various combinations of anti-commuting measurement operators 
can induce unexplored non-trivial phenomena. 
In this work, we shall study a specific example of such measurement-induced phenomenon in measurement-only circuits, being motivated by recent studies \cite{Ippoliti2021,Lavasani2021,Lavasani2021_2,Klocke2022}. 

In this study, a measurement-only circuit with three different types of projective measurements is focused. 
We show that interplay of the specific setting of location of the measurements and suitable combination of 
the commutation relations among the three projective measurements 
induces a topological phase transition, generating SPT states. 
One of the three projective measurements eliminates the disorder projective measurement, which hinders the SPT, and as a result,
the state recovers a specific SPT defined on even sublattice in temporal evolution of the circuit. 
This mechanism seems simple but is definitely interesting as a clear phase transition to sublattice SPT is observed.
After explaining essential ingredients of the mechanism, numerical study shows that this transition belongs to the universality class of the classical 2D percolation. 
Furthermore, we verify that its criticality is universal, independent of a choice of disorder measurements with or without the symmetries of the SPT. 
This numerical result indicates that the criticality of the 2D percolation emerges universally in measurement-only circuit transition regardless 
of the symmetry of operators hindering the SPT, i.e., whether the operators preserve or break the SPT symmetry. 


{\it Circuit setting.---} 
We consider a $L$($=$even integer)-site spin system with open boundary conditions and study measurement-only circuits 
with three different types of projective measurement, $\hat{M}^{a,b,e}_{j}$, which are defined as 
\begin{eqnarray}
{\hat M}^{a}_{j}&=&Z_{2j}X_{2j+1}X_{2j+2}X_{2j+3}Z_{2j+4}\nonumber\\
{\hat M}^{b}_{j}&=&X_{2j}Z_{2j+1}X_{2j+2}Z_{2j+3}\label{Ms}\\
{\hat M}^{e}_{j}&=&X_{2j+1},  \nonumber
\end{eqnarray}
where $j=0,1,\cdots, j^{\alpha}_{max}$ for ${\hat M}^{\alpha}_{j}$ ($\alpha=a,b,e$), and
$j^{a}_{max}=L/2-3$, $j^{b}_{max}=L/2-2$ and $j^e_{max}=L/2-1$.
Each set of the three kinds of operators is a stabilizer set, i.e., 
$[\hat{M}^{\alpha}_{i},\hat{M}^{\alpha}_{j}]=0$ and $(\hat{M}^{\alpha}_j)^2=1$ for $\alpha=a,b,e$. 
The projective measurement of the stabilizer $\{{\hat M}^{a}_j\}$ generates a stable SPT phase 
with the protection symmetry, $Z_2\times Z_2$  \cite{Son2011,Son2012,Bahri2015} or $Z_2\times Z^T_2$ \cite{Verrsen2017,Smith2019}. 
In this work, we focus on the $Z_2\times Z_2$ symmetry, whose generators are $g_{1}\equiv \prod^{L/4}_{i=1}X_{4i}$ and 
$g_{2}\equiv \prod^{L/4}_{i=1}X_{4i+2}$, which commute with $\{\hat{M}^{\alpha}_{\ell}\}$, i.e.,
all of the measurement operators $\{M^{\alpha}_j\}$ respect the $Z_2\times Z_2$ symmetry.
We note that two sets of $\{{\hat M}^{a}_{j}\}$ and $\{\hat{M}^{e}_{j}\}$ commute with each other but, 
some pairs of operators out of $\{\hat{M}^{b}_{j}\}$ and $\{\hat{M}^{e}_{j}\}$ anti-commute with each other. 
We shall show shortly that this property derives important consequences for the dynamics of the system as observed from  the practical update 
in the stabilizer numerics \cite{Gottesman,Aaronson2004}. 

\begin{figure}[t]
\begin{center} 
\includegraphics[width=7.8cm]{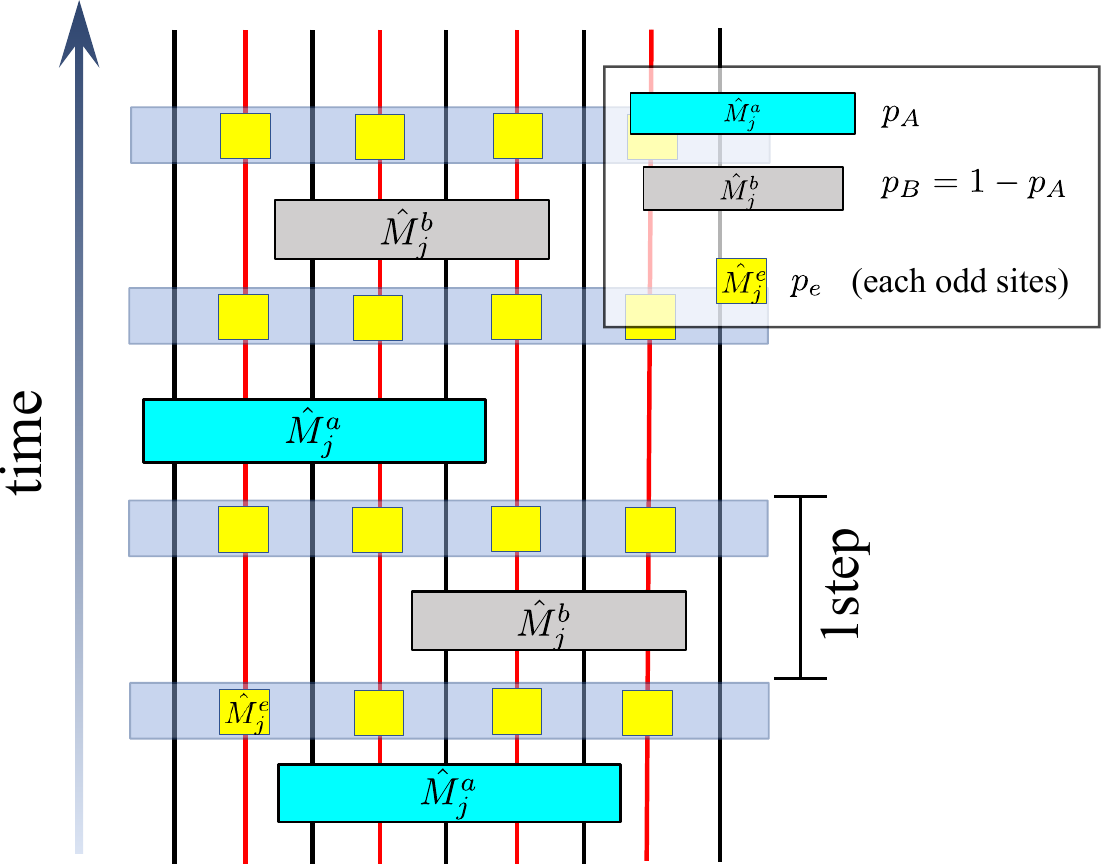}  
\end{center} 
\caption{Schematic figure of measurement only circuit. 
Black and red line represents even and odd site dynamical lines. 
Right blue and gray blocks represent projective measurement $\hat{M}^a_j$ and $\hat{M}^b_j$. 
Yellow blocks represent projective measurement of ${\hat M}^e_j$ applied only on odd sites.}
\label{Fig1}
\end{figure}

The measurement and time-step process in this circuit are as follows: 
In each time step, the following two projective measurements are applied. 
First, we choose $\hat{M}^a$ or $\hat{M}^b$ with probability $p_A$ or $p_B=1-p_A$ and 
choose a measurement site $j$ with uniform probability, and then perform the chosen projective measurement~\cite{Lavasani2021,Klocke2022}. 
Second, we apply the $\hat{M}^e_j$ {\it for each $j$ (odd sites from Eq.~(\ref{Ms})) with the probability} $p_e$, that is, 
$p_e$ is a probability density. 
Please notice the difference between the meaning of the probabilities $p_{A/B}$ and $p_e$.
Single time-step process (the unit of time) consists of the above two successive measurements. 

In addition, we comment that, in the present setup, the projective measurement ${\hat M}^e_j$
can be regarded as error emerging only on odd sites with the probability $p_e$. 
On the other hand, the other two types of measurements $\hat{M}^{a(b)}_j$ can be regarded as a syndrome and a disorder measurement
disturbing the syndrome measurement in error correction literature, respectively.

{\it Intuitive picture of elimination of $\hat{M}^b_j$.---} 
We can observe an essential mechanism for eliminating the disorder measurement ${\hat M}^{b}_j$. 
Assuming that, for simplicity, system is periodic and the initial state in the bulk is prepared by the stabilizer sets of $\{\hat{M}^a_j\}$ and 
$\{\hat{M}^e_j\}$
(See Supplemental Material \cite{Supp}), then we apply a single measurement of the disorder operator at site $j_0$, ${\hat M}^{b}_{j_0}$, 
in the bulk. 
This operation destroys local SPT by $X_{2j_0}Z_{2j_0+1}X_{2j_0+2}Z_{2j_0+3}$.
Then, we further perform the measurement ${\hat M}^e_{j_0}={X_{2j_0+1}}$. 
This process is described by an update flow on a check matrix form in the stabilizer formalism as shown in Refs.~\cite{Gottesman,Aaronson2004};
\begin{eqnarray}
\begin{pmatrix}
\vdots\\
X_1\\
Z_2X_3X_4X_5Z_6\\
X_3\\
Z_4X_5X_6X_7Z_8\\
X_5\\
\vdots\\
\end{pmatrix}
\xrightarrow[]{\text{Measure ${\hat M}^b_1$}}
\begin{pmatrix}
\vdots\\
X_1\\
X_2Z_3X_4Z_5\\
Z_2I_3X_4X_5Z_6\\
Z_4X_5X_6X_7Z_8\\
Z_2X_3X_4I_5Z_6\\
\vdots\\
\end{pmatrix}\nonumber\\
\xrightarrow[]{\text{Measure ${\hat M}^e_1$}}
\begin{pmatrix}
\vdots\\
X_1\\
X_3\\
Z_2I_3X_4X_5Z_6\\
Z_4X_5X_6X_7Z_8\\
Z_2X_3X_4I_5Z_6\\
\vdots\\
\end{pmatrix}
\xrightarrow[]{\text{!\:\:\:}}
\begin{pmatrix}
\vdots\\
X_1\\
Z_2X_3X_4X_5Z_6\\
X_3\\
Z_4X_5X_6X_7Z_8\\
X_5\\
\vdots\\
\end{pmatrix}.
\end{eqnarray}
Here, the finial transformation is only to redefine the stabilizer {\it generators} in basic transformations of check matrices \cite{Nielsen_Chuang}. 
From the above update example, the initial bulk SPT order is protected by eliminating effects of the disorder measurement ${\hat M}^{b}_{j_0}$ by the subsequent
measurement ${\hat M}^{e}_{j_0}$. 
This simply comes from the combination of the commutation relations, $[{\hat M}^{a}_j,{\hat M}^e_j]=0$ and $\{\hat{M}^{a}_j,{\hat M}^e_j\}=0$.
Spatial location of the measurement of ${\hat M}^{e}_{j_0}$ is important in the above process. 

Based on the above observation on the practical process, in the case that the initial state is specified uniquely by all the operators
$\{X_j\}$ ( $j=0,\cdots,L-1$)
and the circuit satisfies the condition such as $p_A<p_B$, 
we expect that (I) for $p_e=0$, the SPT order is not be sustained since the measurement $\hat{M}^b_j$ strongly disturbs it, 
(II) for a sufficiently large $p_{e}$, the measurement ${\hat M}^e_j$ can eliminate the effects of the disorder measurement $\hat{M}^b_j$ and the SPT re-emerges. 
How the SPT is generated (or sustained) by increasing $p_e$ and whether a phase transition takes place at a finite critical probability $p_{e}$ (denoted by $p^c_{e}$)
are nontrivial problem. 
In what follows, we shall clarify these issues by numerical approach.

{\it Observables and numerical approach.---}
The measurement-only circuit can be simulated efficiently by using the algorithm of the stabilizer circuit 
for large system sizes \cite{Gottesman,Aaronson2004}. 
In what follows, we fix $p_A=0.3$ and $p_B=0.7$ and vary $p_e$ as a controllable parameter.
The initial states are prepared by applying the full elements of $\{X_j\}$.

\begin{figure}[t]
\begin{center} 
\includegraphics[width=9cm]{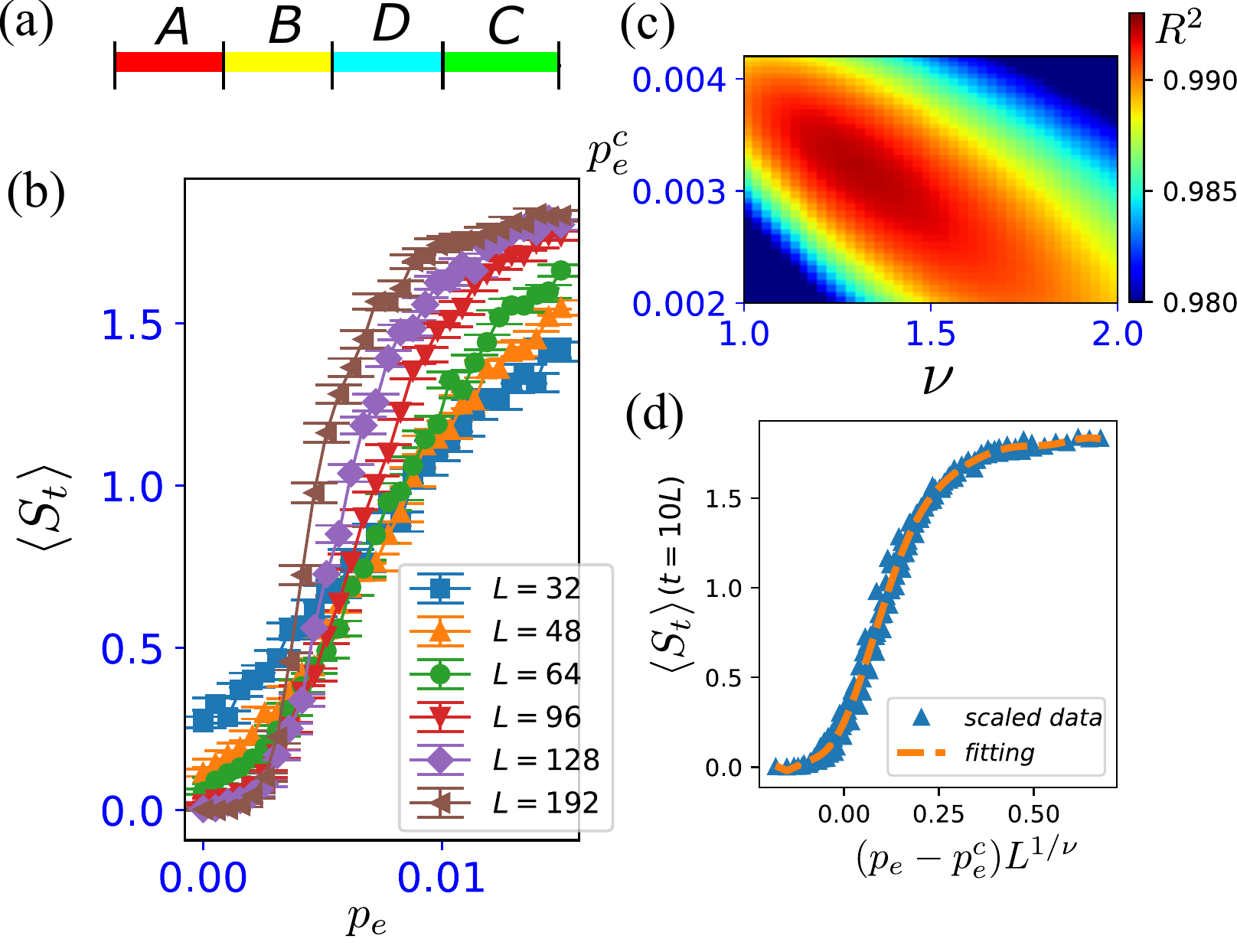}  
\end{center} 
\caption{(a) The system partition for the calculation of the topological entanglement entropy, $S_{t}$. 
Each four subsystem has $L/4$ lattice sites.
(b) Saturation values as increasing $p_e$ for different system sizes. The values are those at $t_N=10L$.
(c) $R^2$ distribution on $\nu$-$p_e$ plane. The optimal values are $R^2=0.992(4)$, $\nu=1.30(0)$ and $p^c_e=0.00315(5)$.  
(d) The scaling fitting function and scaled data set for $\nu=1.30(0)$ and $p^c_e=0.00315(5)$. 
The scaling fitting function is 10-th degree polynomial function. For all data, we set $p_A=0.3$, $p_B=0.7$.}
\label{Fig2}
\end{figure}

To capture dynamics of the system, we observe entanglement behavior of the system. 
Here, in the stabilizer formalism, entanglement entropy for a subsystem $X$ is given by \cite{Fattal2004,Nahum2017}, $S_{X}=n_X- \mathrm{rank}|M_X|$, 
where $n_X$ is system size of the subsystem $X$ and 
$\mathrm{rank}|M_X|$ is the number of linear-independent stabilizers defined within the subsystem $X$ \cite{lin_ent}. 
Then, by partitioning the system into four subsystems as shown in Fig.~\ref{Fig2}(a), we introduce topological entanglement entropy \cite{QI_text,Zeng2016}\\
$$
S_t=S_{AB}+S_{BC}-S_A-S_{ABC}.
$$
System with the exact SPT order has $S_t=2$, which is nothing but the number of the generator of $Z_2\times Z_2$ transformation, 
corresponding to the logical operators of the system under open boundary conditions. 
In our target case with probability $p_A=0.3$, $p_B=0.7$ and $p_e=0$, $Z_2\times Z_2$ SPT state does {\it not} appear as a steady state,
since sufficiently frequent measurement of $\hat{M}^b_j$ destroys the SPT, and then, the saturation value of $S_t$ is much smaller than 2.

For the numerical calculation for the target observables, we employ $800-1000$ different 
measurement patterns for various system sizes and various values of $p_e$, and 
take an ensemble average of saturation value of the observables at $t_N=10L$, where the state is expected to reach a steady state. 

{\it Scaling analysis of phase transition induced by single spatially-fixed error.---}
We calculate averages of the saturation values of $S_t$ denoted by $\langle S_t\rangle$ 
as a function of $p_e$ for various system sizes. 
The results are shown in Fig.~\ref{Fig2}(b). 
The data indicate the existence of a phase transition since all data obtained for various system sizes cross 
with each other at a single point.

To detect the phase transition point and examine its criticality, we employ the finite-size scaling (FSS) analysis for 
$\langle S_{t}\rangle$. 
Here, we assume the following scaling ansatz,
$$
\langle S_{t}\rangle (p_e, L)=\Psi((p_e-p^c_e)L^\nu),
$$
where $\Psi$ is a scaling function, $\nu$ is the critical exponent of a typical length and 
$p_e^c$ is the critical measurement probability for $\hat{M}^e_j$. 
In the FSS, we determine the scaling function $\Psi$, by using the fitting methods of the FSS;
the fitting curve for the scaling function is obtained via a 10-th order polynomial function with the best optimal coefficients 
for various values of $p_e$ and $\nu$, and 
then the coefficient of determination $R^2$ is estimated.
The $R^2$ quantifies how accurately the values of $\langle S_t\rangle$ with different system sizes collapse to a single curve.
Similar procedure was previously employed to study phase-transition properties in similar measurement-only circuits 
and obtained reliable results \cite{Lavasani2021,Klocke2022}.\\ 

The FSS result is displayed in Figs.~\ref{Fig2}(c) and \ref{Fig2}(d), obtained from the $L=48-192$ data. 
The critical transition rate is estimated as $p^c_e=0.00315(5)$ and the exponent $\nu=1.30(0)$. 
The obtained value of $\nu$ is significantly close to that of the 2D percolation $\nu=4/3$, 
which has been reported in similar SPT phase transitions in measurement-only circuits \cite{Lavasani2021,Klocke2022}.
This result indicates that the present circuit including elimination process of the disorder projective measurement 
$\{\hat{M}^b_j\}$ belongs to the same universality class of these systems. 

In the above, we studied the topological entanglement entropy $S_t$ to observe the SPT-ordered states. 
To verify and support the existence of the SPT in the bulk, we measure the {\it even-site
only} Edward-Anderson type of string topological order parameter (STO)~\cite{Lavasani2021}, 
which is defined as follows, 
$$
|{\rm STO}|^2 = |\langle \psi(t) | G^s(i_0,j_0) |\psi(t)\rangle|^2, 
$$
where $|\psi(t)\rangle$ is a unique stabilizer state at a time $t$, i.e., $s^{\ell}(t)|\psi(t)\rangle=|\psi(t)\rangle$
[$\{s^{\ell}\}$ is the stabilizer set at $t$], and 
$$
G^s(i_0,j_0)=Z_{2i_0} Y_{2(i_0+1)}\biggl(\prod^{j_0-2}_{i_0+2}X_{2k}\biggr)Y_{2(j_0-1)}Z_{2j_0}.
$$
The practical calculation scheme in numerics is explained in Supplemental Material \cite{Supp}. 
Figure~\ref{Fig3} shows the result of the averaged saturation values of STO, $\langle|{\rm STO}|^2\rangle$,
sampled at $t_N=10L$ as a function of $p_e$ for various system sizes. 
The numerical results indicate that the averaged STO for large systems with large $p_e$ exhibit sufficiently large values, i.e., 
giving a clear evidence for the existence of the bulk SPT order and phase transition in the bulk, 
in contrast to the claim in a similar calculation in Ref.\cite{Lavasani2021}.

\begin{figure}[t]
\begin{center} 
\includegraphics[width=7cm]{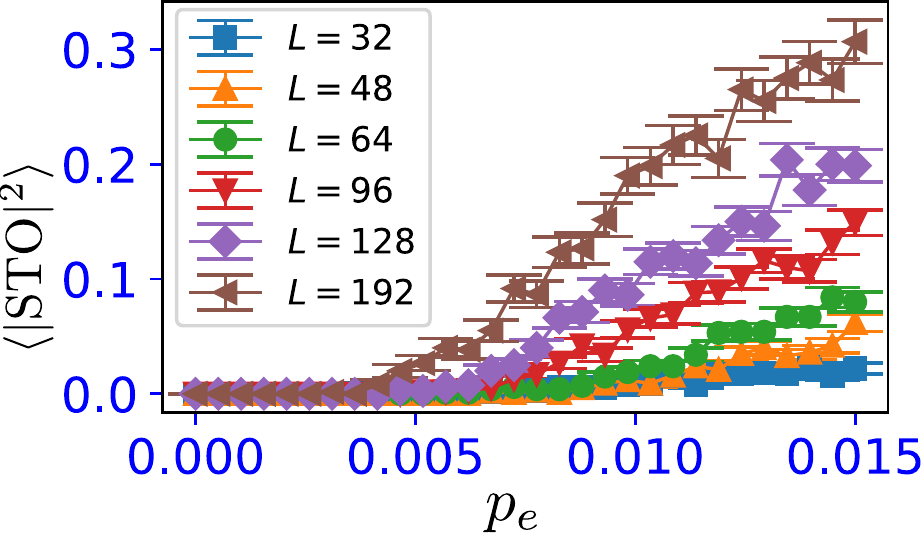}  
\end{center} 
\caption{Behavior of saturation values of the string topological order (STO) in the bulk.
For all data, we set $p_A=0.3$ and $p_B=0.7$, $i_0=2$ and $j_0=L/2-3$ for the calculation of the STO.}
\label{Fig3}
\end{figure}

We observe that the STO is an increasing function of $p_e$ for all system sizes, 
which is a signal of the existence of the SPT in the bulk defined on even sites. 
We further carry out the FSS for the STO and the obtained results are shown in Supplemental Material \cite{Supp}.
The estimated values  of $p^c_e$ and $\nu$ are fairly close to those obtained by the FSS of $S_t$.
We expect that the FSS of $\langle S_t\rangle$ is more reliable than that of the STO as the $R^2$-analysis indicates.

{\it Symmetry breaking type measurement ${\hat M}^d_j$.---}
We investigate effects by a different type of measurement operator instead of $\{{\hat M}^b_j\}$. 
Here, as a specific example, we consider
\begin{eqnarray}
{\hat M}^{d}_{j}&=&Z_{2j}Z_{2j+1}Z_{2j+2}Z_{2j+3}Z_{2j+4},
\end{eqnarray}
where $j=0,1,\cdots, j^{b}_{max}$ and  $j^{b}_{max}=L/2-3$.
$\{\hat{M}^{d}_{j}\}$ is a stabilizer set, i.e., $[M^{d}_{i},M^{d}_j]=0$, and $(M^{d}_j)^2=1$.
They satisfy the same commutation relations with the former set of measurement operators, i.e.,
the two sets of $\{{\hat M}^{a}_{j}\}$ and $\{\hat{M}^{e}_{j}\}$ commute with each other, but
certain pairs out of two sets of $\{\hat{M}^{d}_{j}\}$ and $\{\hat{M}^{e}_{j}\}$ anti-commute. 

\begin{figure}[t]
\begin{center} 
\includegraphics[width=9cm]{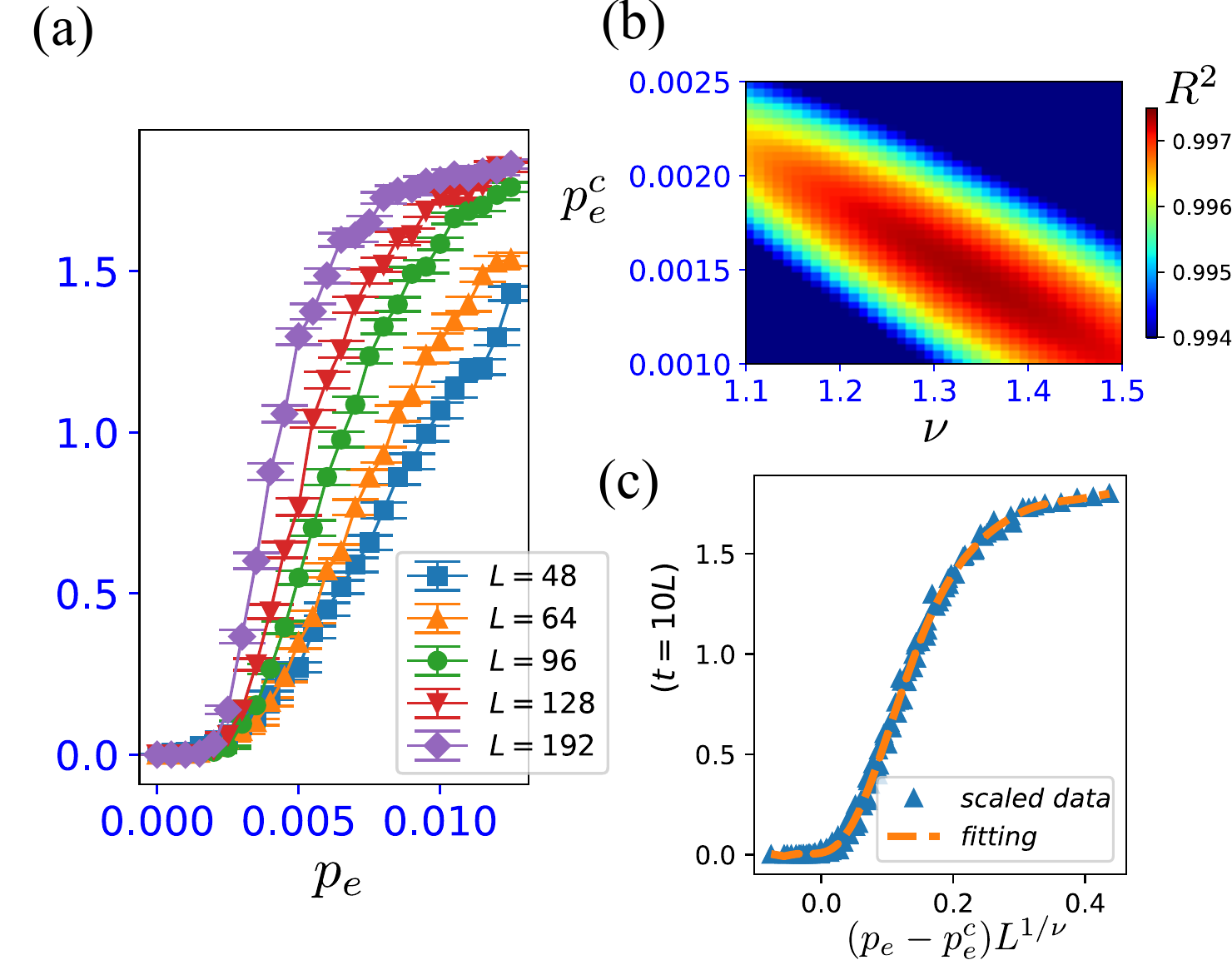}  
\end{center} 
\caption{
(a) Saturation values as a function of $p_e$ for various system sizes. The values are those at $t_N=10L$.
(b) $R^2$ distribution on $\nu$-$p_e$ plane. 
The optimal values are $R^2=0.997(8)$, $\nu=1.36(2)$ and $p^c_e=0.00152(2)$.  
(c) The scaling fitting function and scaled data set for $\nu=1.36(2)$ and $p^c_e=0.00152(2)$. The scaling fitting function is 10-th degree polynomial function. 
For all data, we set $p_A=0.3$, $p_B=0.7$.}
\label{Fig4}
\end{figure}

The $\{\hat{M}^{d}_{j}\}$ projective measurement operators, however, do {\it not} respect the $Z_2\times Z_2$ symmetry. 
Thus, these projective operators strongly suppress the emergence of 
the $Z_2\times Z_2$ SPT phase even for small $p_D=1-p_A$ and $p_e=0$ 
since the topological entanglement entropy $S_t$ vanishes due to the symmetry breaking effects.
Under this situation, we carry out similar calculations as in Fig.~\ref{Fig2}. 
The results are shown in Fig.~\ref{Fig4}. 

We also observe an elimination transition similar to the former case as shown in Fig.~\ref{Fig4}(a) with 
the FSS analysis summerized in Figs.~\ref{Fig4}(b) and \ref{Fig4}(c). 
From the FSS of $L=48-192$ data, the critical transition rate and exponent are estimated as $p^c_e=0.00152(2)$ and $\nu=1.36(2)$. 
This value of $\nu$ is substantially close to the 2D percolation $\nu=4/3$ as in the former case.
Therefore, the numerics of this case indicates that even for finite probability rate of $\{\hat{M}^{d}_{j}\}$ that breaks the $Z_2\times Z_2$ symmetry, 
the symmetry breaking effect is also eliminated by the projective measurement $\{\hat{M}^e_j\}$ for sufficiently large $p_e$, 
and the $Z_2\times Z_2$ SPT order appears as a steady state. 
Thus, the numerical result in this work indicates that the 2D percolation picture is fairly universal at the criticality of phase transitions 
generating SPT steady states.

{\it Conclusion.---}
Specific spatial setting and combination of commutation relations among three measurement operators induce a novel type of topological phase transition 
in a measurement-only circuit, where the steady state has a SPT order defined on sublattice only. 
This SPT is recovered by eliminating a  projective measurement suppressing the SPT order. 
This elimination process is simple, but spatially well-located measurement induces a clear measurement-only phase transition, not a crossover.
The detailed analysis of criticality exhibits robustness of the 2D percolation universality class at the criticality of phase transitions generating of SPT steady states. 
We expect that various spatially-tuned measurements with various 
kinds of projective operators induce a rich non-trivial phase transition in measurement-only circuits.

{\it Acknowledgements.---}
This work is supported by JSPS KAKEN-HI Grant Number JP21K13849 (Y.K.). 

\bibliography{main_bib_test}

\begin{thebibliography}{45}
\bibitem [{lin()}]{lin_ent}%
  \BibitemOpen
  \href@noop {} {\ }\bibinfo {note} {$|M_X|$ is a portion of the check matrix,
  which is composed of vertically stacked vector representations of stabilizer
  operators represented by each component of the Pauli operator
  \cite{Nielsen_Chuang}. We calculated the rank of $|M_X|$ by a binary Gaussian
  elimination.}\BibitemShut {Stop}%
\end{thebibliography}

\clearpage
\appendix
\section*{Supplemental material}


\section*{A: Tractable update behavior to SPT for small size system in the limit $p_A=1$ and $p_e = 1$.}
Let us consider $L=12$ system with open boundary conditions and start with the initial state specified by the full elements of 
$\{X_j\}$ ($j=0,\cdots, 11$).
The circuit of the limiting $p_A=1$ case dynamically creates an ideal $Z_2\times Z_2$ SPT order defined only on the even sublattice,
even though the measurement of $\hat{M}^e_j$ does not affects the time evolution of the circuit. 
By applying measurement of the operator $\hat{M}^{a}_j$ for a uniform random $j$ many times, the state is obliged to reach the following steady state 
\begin{eqnarray}
\begin{pmatrix}
X_0\\
X_1\\
X_2\\
X_3\\
X_4\\
X_5\\
X_6\\
X_7\\
X_8\\
X_9\\
X_{10}\\
X_{11}\\
\end{pmatrix}
\xrightarrow[]{\text{Measure all ${\hat M}^a_j$}}
\begin{pmatrix}
{\hat M}^a_0 \\
X_2\\
{\hat M}^a_1\\
X_3\\
{\hat M}^a_2\\
X_5\\
{\hat M}^a_3\\
X_7\\
{\hat M}^a_4\\
X_0X_4X_8\\
X_2X_6X_{10}\\
X_{11}\\
\end{pmatrix}\nonumber\\
\xrightarrow[]{\text{Pick up even sites}}
\begin{pmatrix}
Z_0I_1X_2I_3Z_4\\
Z_2I_3X_4I_5Z_6\\
Z_4I_5X_6I_7Z_8\\
Z_6I_7X_8I_9Z_{10}\\
X_0X_4X_8\\
X_2X_6X_{10}\\
\end{pmatrix}.
\end{eqnarray}
Here, we focus on the even sites and recognize that the linearly-independent stabilizers are the ones of the $Z_2\times Z_2$ SPT and 
the corresponding logical operators. 
Thus, on the even sublattice, $Z_2\times Z_2$ SPT state appears. 
Also, note that the stabilizers of the SPT and 
logical operators are not affected by the projective measurement of $\hat{M}^e_j$ since these measurements are performed on odd sites.

\section*{B: Computation of string topological order and scaling analysis}
We briefly explain how to calculate the STO in the numerics. 
The density matrix of the system (a pure state) is written by 
$$
\rho(t)=\prod^{L-1}_{\ell=0}\biggr( \frac{1+s^{\ell}(t)}{2}\biggl),
$$
where $s^{\ell}(t)$ is updated stabilizers at a time $t$. 
We calculate the Edward-Anderson grass-like string order
$$
|\langle {\rm STO}\rangle |^2 = |\langle \psi(t) | G^s |\psi(t)\rangle|^2, 
$$
where $|\psi(t)\rangle$ is a unique stabilizer state at a time $t$, 
$s^{\ell}(t)|\psi(t)\rangle=|\psi(t)\rangle$ for ${}^{\forall}\ell$,
and
$$
G^s(i_0,j_0)=Z_{2i_0} Y_{2(i_0+1)}\biggl(\prod^{j_0-2}_{i_0+2}X_{2k}\biggr)Y_{2(j_0-1)}Z_{2j_0}.
$$

The STO is calculated in the stabilizer formalism as $G^s(i_0,j_0)$ is only written by Pauli string without imaginary factor $i$ and $(G^{s})^2=1$. 
Each stabilizer $s^{\ell}(t)$ commutes or anti-commutes with $G^s$ at ${}^{\forall} t$, 
$s^{\ell}(t)G^s=\alpha^{\ell}_{\pm}G^s s^{\ell}(t)$ with $\alpha^{\ell}_{\pm}=\pm 1$. 
The STO is reduced to a simple form
\begin{eqnarray}
&&| {\rm STO} |^2 = \frac{1}{2^L}\langle \psi(t) | G^s \biggl(\prod^{L}_{\ell=1}(1+s^{\ell}(t))\biggr)G^s |\psi(t)\rangle\nonumber\\
&&=\frac{1}{2^{L}}\prod^{L}_{\ell=1}\langle \psi(t)|(1+\alpha^\ell_{\pm}s^{\ell}(t))|\psi(t)\rangle\nonumber\\
&&=\frac{1}{2^{L}}\prod^{L}_{\ell=1}(1+\alpha^\ell_{\pm}), \nonumber
\end{eqnarray}
where we used $G^s(1+s^{\ell}(t))G^s=(1+G^ss^{\ell}(t)G^s)=(1+\alpha^{\ell}_{\pm}s^{\ell}(t))$ and $s^{\ell}(t)|\psi(t)\rangle =|\psi(t)\rangle$ at ${}^{\forall} t$. 
For the ideal $Z_2\times Z_2$ SPT phase, due to $\alpha_{\pm}^{\ell}=1$ for ${}^{\forall} \ell$, $| {\rm STO} |^2=1$ 
while for no $Z_2\times Z_2$ SPT phase, strictly $| {\rm STO} |^2=0$ due to due to $\alpha_{\pm}^{\ell}=-1$ for ${}^{\forall} \ell$.

\begin{figure}[b]
\begin{center} 
\vspace{0.5cm}
\includegraphics[width=8cm]{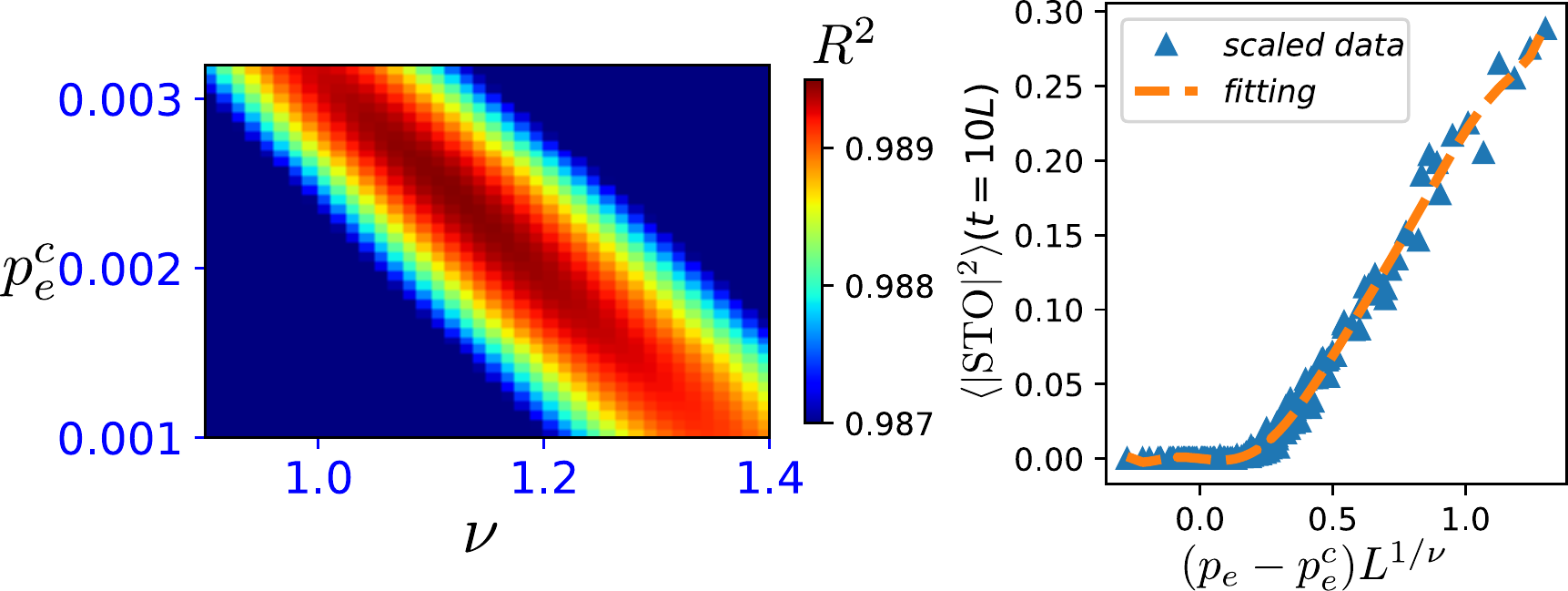}  
\end{center} 
\caption{Scaling analysis of the STO: 
(Left) $R^2$ distribution on $\nu$-$p_e$ plane. The optimal values are $R^2=0.989(5)$, $\nu=1.113(5)$ and $p^c_e=0.00243(1)$.  
(Right) The scaling fitting function and scaled data set for $\nu=1.113(5)$ and $p^c_e=0.00243(1)$. 
The scaling fitting function is 10-th degree polynomial function. For all data, we set $p_A=0.3$, $p_B=0.7$.}
\label{Fig5}
\end{figure}

\section*{C: Scaling analysis of the string order}

We show the finite-size scaling (FSS) analysis for the STO defined in the main text. 
In the FSS, we use the averaged saturation values of the STO, $\langle |\rm STO|^2\rangle$, in the same parameter setting ($p_A=0.3$ and $p_B=0.7$)
and the saturation values are taken at $t=10L$, as shown in Fig.~\ref{Fig3} in the main text. 
To detect phase transition point of $p_e$ and examine its criticality from the data of the STO in the bulk, 
we carried out the same approach to the main manuscript. 
Here, we assume a scaling function of the same form to the topological entanglement entropy,
$$
\langle |{\rm STO}|^2\rangle (p_e, L)=\Phi((p_e-p^c_e)L^\nu),
$$ 
where $\Phi$ is a scaling function and $\nu$ is the critical exponent and $p_e^c$ is the critical measurement probability. 

The FSS result is summerized in Fig.~\ref{Fig5}. 
From the fitting by using $L=48-192$ data, 
we obtained $R^2=0.989(5)$ as the best optical. 
The optimal critical transition rate is estimated as $p^c_e=0.00243(1)$ and the optimal critical exponent also $\nu=1.113(5)$. 
Compared to the FSS results of the topological entanglement entropy in the main text, 
the values of $p^c_e$ and $\nu$ are slightly smaller than those of the topological entanglement entropy, 
but the estimated value of $\nu$ is close to the 2D percolation $\nu=4/3$. 
Since the value of $R^2$ in the FSS of the STO is slightly smaller that that of the topological entanglement entropy, 
the results of the FSS of the topological entanglement entropy is more reliable.

\bigskip

\end{document}